\documentclass[english,acs,jce,reprint,twocolumn]{revtex4}
\usepackage[T1]{fontenc}
\usepackage{textcomp}
\usepackage{bm}
\usepackage{amstext}
\usepackage{graphicx}
\usepackage{esint}

\makeatletter

\newcommand{\lyxmathsym}[1]{\ifmmode\begingroup\def\b@ld{bold}
  \text{\ifx\math@version\b@ld\bfseries\fi#1}\endgroup\else#1\fi}

\@ifundefined{textcolor}{}
{%
 \definecolor{BLACK}{gray}{0}
 \definecolor{WHITE}{gray}{1}
 \definecolor{RED}{rgb}{1,0,0}
 \definecolor{GREEN}{rgb}{0,1,0}
 \definecolor{BLUE}{rgb}{0,0,1}
 \definecolor{CYAN}{cmyk}{1,0,0,0}
 \definecolor{MAGENTA}{cmyk}{0,1,0,0}
 \definecolor{YELLOW}{cmyk}{0,0,1,0}
}

\makeatother

\usepackage{babel}
\begin{document}

\title{Introduction to Classical Density Functional Theory by a Computational Experiment}

\author{Guillaume Jeanmairet$^{1,2,3,}$\footnote{guillaume.jeanmairet@ens.fr}
}
\author{Nicolas Levy$^{1,2}$}
\author{Maximilien Levesque$^{1,2,3,}$\footnote{maximilien.levesque@ens.fr}
}
\author{Daniel Borgis$^{1,2,3}$}

\affiliation{$^1$\'Ecole Normale Sup\'erieure-PSL Research University, D\'epartement de Chimie, 24, rue Lhomond, 75005 Paris, France.\\
$^2$Sorbonne Universit\'es, UPMC Univ Paris 06, PASTEUR, F-75005, Paris, France.\\
$^3$CNRS, UMR 8640 PASTEUR, F-75005, Paris, France.
}

\begin{abstract}
We propose an in-silico experiment to introduce classical density functional theory (cDFT). Density functional theories, whether quantum or classical, rely on abstract concepts that are non-intuitive. However, they are at the heart of powerful tools and active fields of research in both physics and chemistry. They led to the 1998 Nobel Prize in chemistry. DFT is illustrated here in its most simple and yet physically relevant form: the classical density functional theory of an ideal fluid of classical particles. For illustration purpose, it is applied to the prediction of the molecular structure of liquid neon. The numerical experiment proposed therein is built around the writing of a cDFT code by students in Mathematica. Students thus have to deal with (\emph{i}) the cDFT theory, (\emph{i}i) some basic concepts of statistical mechanics of simple fluids, (\emph{iii}) functional minimization, and (\emph{iv}) a useful functional programming language. This computational experiment is proposed during a molecular simulation class, but may also be of interest in a quantum chemistry class to illustrate electronic DFT, if one highlights the analogies between the quantum and classical DFTs.\end{abstract}
\maketitle
Is it possible to study and extract all interesting information of a quantum system without considering explicitly, i.e., individually, all the electrons? Hohenberg, Kohn and Sham answered yes to these questions when they invented the electronic density functional theory (DFT). In their seminal papers, published in the middle of the 1960's, they show that the knowledge of the number of electrons per volume unit, the electronic density, is a sufficient quantity to extract all the properties of any system. They also show that finding this density is an easy process: it is the density that minimizes a free energy functional (a function of a function) of the density itself.\cite{kohn_self-consistent_1965,hohenberg_inhomogeneous_1964}
Thanks to its high computational efficiency with respect to theories based on electronic wavefunctions, electronic density functional theory has imposed as a major theoretical tool to treat quantum mechanics problems in academic labs.\cite{dreizler_introduction_1990,argaman_density_2000} As a consequence, it takes a prominent part in the current quantum chemistry courses (for instance, half of the hours dedicated to theoretical chemistry at \'Ecole Normale Sup\'erieure, Paris, France). 
The determination of the energetic and structural equilibrium properties of classical systems is also an important problem. Often, it is addressed in statistical mechanics or molecular simulation courses for undergraduate students. During these courses, Molecular Dynamics (MD) or Monte-Carlo simulations (MC)\cite{frenkel_understanding_2002} are often mentioned. However, these methods are most often introduced theoretically because writing or using MD or MC codes is time-consuming. A classical density functional theory framework exists, that is similar to electronic DFT, for systems of classical particles. The electronic probability density is at the heart of the electronic DFT, while it is the number of molecules per unit volume in classical DFT.
Another interest of liquid state theories like cDFT in molecular simulation courses is to illustrate the problem of computational costs that often are a limitation of everyday research. cDFT\cite{evans_nature_1979,mermin_thermal_1965,moore_teaching_1998} is theoretically as accurate as MC and MD, while reducing computational costs by two to three orders of magnitude. 

This article discusses a 2 hours in-silico experiment for at most twelve students. It is part of a theoretical chemistry course for chemists or physicists with a minor in chemistry. Students do not need to be familiar with programming nor functional minimization, but prior introduction to DFT and functional analysis is required. To illustrate the use of cDFT, they are asked to write a Mathematica code.\cite{mathematica9} Each student should thus have a computer with Mathematica version 9 or later. This language is chosen because it has both strong symbolic and numeric capabilities and because its functional programming paradigm that allows students not to focus on programming but on algorithms. An example Mathematica notebook is given in Supporting Information to guide the professor. It is not intended to be given to the students before the experimental session. The professor should let students who are familiar with Mathematica to work as independently as possible. Students without prior knowledge of Mathematica should be guided almost in real time: everyone should code together using on-screen projection of the professor's Mathematica notebook. Since the students write the program at the same time as the professor, they can play with the Mathematica documentation, the functions in use and the plots. The professor could give the program found in Supporting Information after the session, in order to let the students have a correct and well-written version of the program for future reference.
The structure of this article is the following: first, the classical density functional theory is introduced quickly. This theoretical part gives rise to a discussion about functionals and their minimization, which often are unfamiliar notions to undergraduate students. Secondly, the 2-hours practical computational exercice discussed above is proposed: Students investigate a model of fluid neon. Students are invited to solve this problem by writing a Mathematica program that applies the previously introduced theoretical framework. This program should minimize the Helmholtz free energy functional both analytically and numerically. Results are then discussed and compared to experimental neutron scattering data.\cite{graaf_structure_2003} Third, the strengths and weaknesses of the theory and approximations made during the course are pointed out, before concluding.

\section{Theory}

Classical density functional theory (cDFT) is the classical analogue of electronic DFT introduced by Kohn (Nobel prize in 1998\cite{Nobel98}) together with his collaborators Hohenberg and Sham.\cite{kohn_self-consistent_1965,hohenberg_inhomogeneous_1964} Electronic DFT is a well known theory, for which good reviews can easily be found.\cite{argaman_density_2000,parr_density_1983} It is based on (i) the rewriting of the Hamiltonian of a system of $N$ electrons in an external field (e.g., the electrostatic potential induced by the nuclei), as a unique functional of the electronic density, and (ii) the fact that the electronic density that minimizes the functional is the density of the ground state, the state of lowest energy. It has been proved\cite{evans_nature_1979,mermin_thermal_1965} that these results are also true for classical systems: the minimization of the Helmholtz free energy ${\cal F}$, that is a unique functional of the classical solvent density $\rho(\bm{r})$, with respect to this density, gives access to the equilibrium density $\rho_{eq}(\bm{r})$. This equilibrium density must be understood as the average quantity of solvent molecules at each position $\bm{r}$ in the presence of an external perturbation, e.g., a solid surface or any solute. Students are asked to think about the fact that ${\cal F}$ is a function of $\rho$ that is itself a function of the position $\bm{r}$.  is thus called a functional and shall be written . Rigorously,\cite{hansen_theory_2006} the free energy functional can be decomposed as in equation (1). 

\begin{equation}
{\cal F}[\rho(\bm{r})]={\cal F}_{id}[\rho(\bm{r})]+{\cal F}_{exc}[\rho(\bm{r})]+{\cal F}_{ext}[\rho(\bm{r})]-\mu\int\rho\left(\mathbf{r}\right)d\mathbf{r},\label{eq:Fdecompose}
\end{equation}
with,
\begin{equation}
{\cal F}_{id}[\rho(\bm{r})] =  k_{B}T\int\rho(\bm{r})(\ln\left[\Lambda^{3}\rho(\bm{r})\right]-1))d\bm{r},\label{eq:Fid}
\end{equation}
and,
\begin{equation}
{\cal F}_{ext}[\rho(\bm{r})]  =  \int v_{ext}(\bm{r})\rho(\bm{r})d\bm{r},\label{eq:Fext}
\end{equation}

Where $\mu$ is the chemical potential, $k_B$ is the Boltzmann constant, and $T$ the temperature in Kelvin. The inverse of the thermal energy,  $\beta=(k_BT)^{-1}$ is used later in this article. $\Lambda=\sqrt{\frac{\beta h^2}{2\pi m}}$ is the thermal wavelength, $h$ is the Planck constant and $m$ is the mass of one solvent molecule. In equations (2)-(3), students are shown that the scalars ${\cal F}_{id}[\rho(\bm{r})]$ and ${\cal F}_{ext}[\rho(\bm{r})]$ depend of all the values taken by $\rho$ at all positions $\bm{r} $.
${\cal F}_{id}[\rho(\bm{r})] $ is the ideal term of a non-interacting (ideal) fluid. It is purely entropic and its expression, given in equation (2), is exact. ${\cal F}_{ext}[\rho(\bm{r})]$ is the term due to an external potential, $v_{ext}(\bm{r})$. It is the perturbation of the solvent. It can be, for instance, the interaction of the solvent with a molecular solute or an interface. ${\cal F}_{exc}[\rho(\bm{r})]$ is the excess term, due to solvent-solvent interactions. Exact expression for this term is, in the general case, still unknown. Finding the best approximation for it, especially for water, is one intense field of research that goes far beyond this introduction.\cite{jeanmairet_molecular_2013} Note that this excess term is the equivalent of the exchange correlation term in electronic DFT: it is approximated. In this first approach of cDFT, it will be neglected. Interested students may refer to reference \cite{hansen_theory_2006} for a review on how to tackle this term.
Given an excess term and an external potential, it is possible to minimize the functional, i.e., to find the solvent density $\rho_{eq}(\bm{r})$ such that ${\cal F}$ is minimum. Once this minimization is done, the equilibrium density, and thus all the structural properties are known.
In order to get used to this functional approach, students are now asked to minimize both numerically and analytically the simplest form for ${\cal F}$. They will then be able to deduce the structural properties of a model of liquid neon.\cite{graaf_structure_2003}

\section{Student exercise and computational method}
In order to limit this in-silico experiment to 2 hours, the minimization will be carried out in one dimension only, reducing the computational and programming efforts. The spatial coordinates $\bm{r}$, are thus discretized on a regular grid of $i$ nodes: $\{r_i\}$. One objective is to treat a case that makes the students able to compare the analytical and numerical minimizations. To further simplify the exercise, the problematic excess term is omitted in the following. This approximation decreases the computational cost of the numerical minimization of the functional by one order of magnitude. Indeed the ideal and the external term require a simple integration over the spatial coordinates $\bm{r}$, as can be seen in equations (2)-(3). The computational cost of such integration scales as ${\cal O}(N)$, where $N$ is the number of grid nodes used to discretize space. The excess term would have required, at least, a double integration on space, and consequently scales at best as ${\cal O}(N^2)$. Furthermore, the excess term makes the analytical functional minimization much more complex, if not impossible. It should be clear at this point that the fluid has no correlations: it is ideal. It is a good opportunity to discuss how the balance between complexity and computational cost is at the heart of scientific research that relies on theory and/or computation. 
First, the students are requested to find, analytically, the functional derivative of the free energy. It reads:

\begin{equation}
\frac{\delta{\cal F}[\rho(\bm{r})]}{\delta\rho(\bm{r})}=\beta^{-1}\ln\left[\Lambda^{3}\rho(\bm{r})\right]+v_{ext}(\mathbf{r})-\mu,\label{eq:dF/drho=00003D0-1}
\end{equation}

Special attention should be given to this step of functional derivation, as it is often difficult for students. Because DFT is a variational principle, the solution is the density that nullifies equation (4), that is:

\begin{equation}
\rho_{eq}(\bm{r})=\frac{\exp\left(\beta\mu\right)}{\Lambda^{3}}\exp\left(-\beta v_{ext}(\bm{r})\right),\label{eq:analytical_res}
\end{equation}
where the chemical potential, $\mu$, is chosen so that $\exp(\beta\mu)/\Lambda^3=\rho_b$, the homogeneous, bulk, solvent density. With this choice, the solvent is homogeneous when the perturbation vanishes. On the contrary, where the perturbation is very repulsive ($v_{ext}\gg k_BT$), the equilibrium density tends toward zero.
Secondly, in order to illustrate the powerful variational principle at the heart of DFTs and to be able to extend this course to more complex cases, students are requested to minimize the free energy numerically. For this purpose, they are asked to discretize equations (2)-(3):

\begin{equation}
{\cal F}_{id}[\rho(\{r_{i}\})]  =  k_{B}T\sum_{i}\Delta_{i}\rho(r_{i})\left[\ln\left(\Lambda^{3}\rho(r_{i})\right)-1\right],\label{eq:fid_discret}
\end{equation}

\begin{equation}
{\cal F}_{ext}[\{\rho(\{r_{i}\})]  =  \sum_{i}\Delta_{i}\rho(r_{i})v_{ext}(r_{i}),\label{eq:fext_discret}
\end{equation}

where $\Delta_i$ is the element of integration, that is in our one-dimensional case the distance between two radial nodes.
To minimize the functional, they need to chose an initial density: naively, students often propose the homogeneous solvent density $\rho_b$, which is fine. Then, they are asked to find and plot the equilibrium density profiles corresponding to two different external potentials: a hard wall and a repulsive field of Gaussian shape.    
Third, in order to compare with a physical system, we propose to focus on liquid neon. Its bulk density is 0.033 molecules/\AA$^3$ at 35.05 K. Such a system was studied experimentally by neutron scattering by de Graaf and Mozer.\cite{graaf_structure_2003} They reported the radial distribution function (rdf) of liquid neon, which is the probability to find a Ne atom at distance $r$ of another Ne, normalized by the bulk density. It is related to the density at a distance $r$ of a reference solvent molecule by

\begin{equation}
g(r)=\frac{\rho(r)}{\rho_{b}}.\label{eq:rdf}
\end{equation}

As no excess term has been included, there is no Ne-Ne interaction: it is an ideal liquid. The most common and widely used interatomic potential for Van der Waals interactions is the Lennard-Jones potential: 

\begin{equation}
v_{ext}(r)=4\epsilon\left[\left(\frac{\sigma}{r}\right)^{12}-\left(\frac{\sigma}{r}\right)^{6}\right],\label{eq:lj}
\end{equation}

with parameters $\epsilon$=1.6 kJ/mol and $\sigma$=2.6 \AA~for Ne.

\section{Results and Discussion}

\begin{figure}
\centering{}\includegraphics[width=0.5\textwidth]{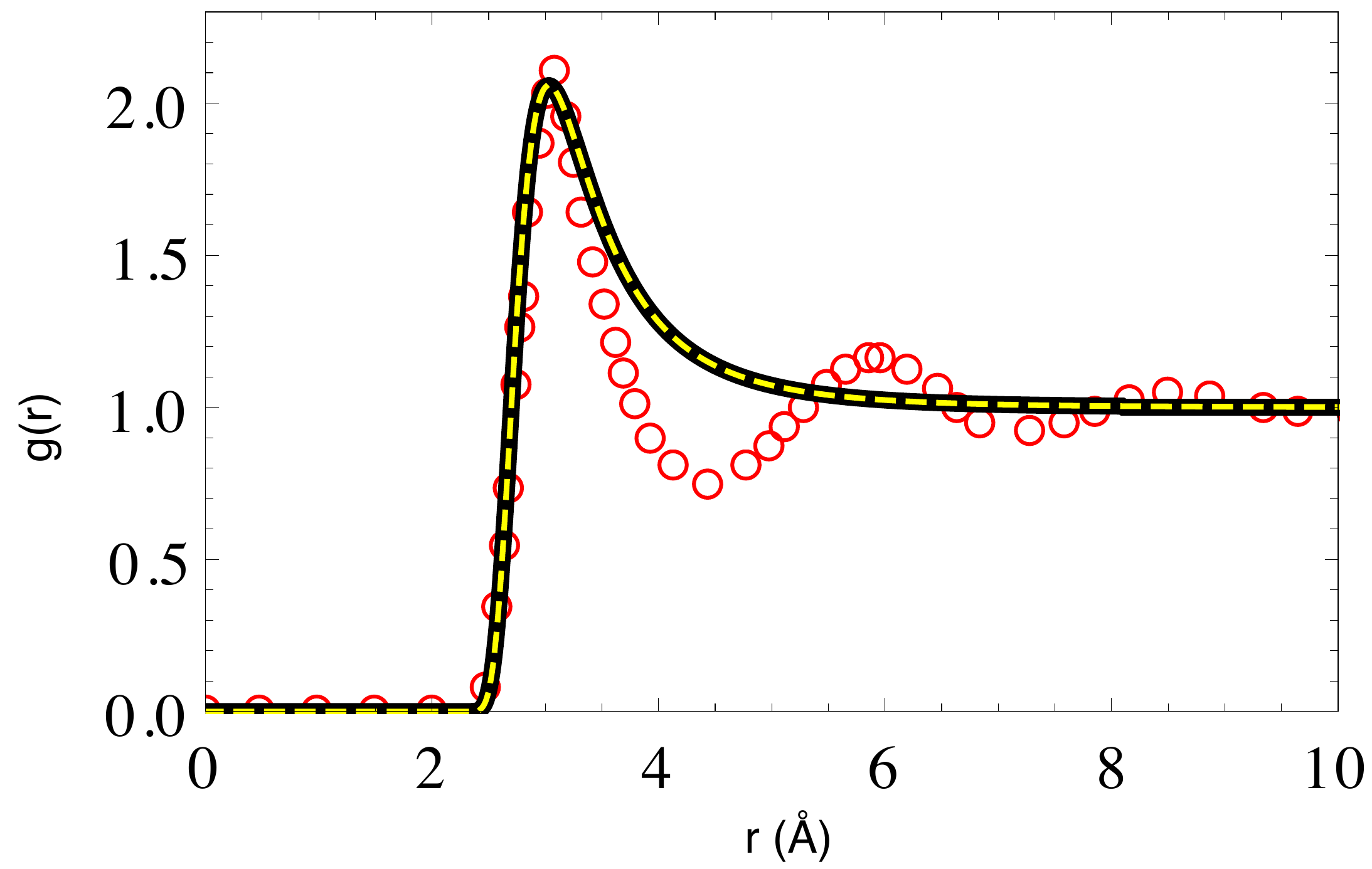}\caption{Radial distribution function for liquid neon. Experimental results by de Graaf and Mozer
\cite{graaf_structure_2003}
are shown in red
circles. The numerical and analytical cDFT results
are shown in black and yellow dashed lines, respectively. \label{fig:Radial-distribution-function}}
\end{figure}
Numerical minimization of the free energy $\cal F$ for our model neon results in the equilibrium solvent density, $\rho(\bm{r})$, from which the rdf is computed with equation (8). It is plotted in full black line in Figure 1. It is also compared to the radial distribution function obtained analytically from equation (5), in yellow dashed lines.
It can be noticed that the radial distribution profiles obtained by analytical and numerical minimizations are identical. Students can be convinced that numerical methods can reproduce exact results predicted by theory, which justifies the use of numerical methods when problems cannot be tackled analytically.
Theoretical predictions and the experimental rdf for liquid neon, extracted from reference 10, are compared in Figure 1. The simple model studied here is able to reproduce the good position for the first peak in the rdf. This peak corresponds to the position of the molecules of the first solvation shell. However, there is no other extrema in the cDFT results because our model does not contain any solvent-solvent interactions. The second peak must be understood as a response to the presence of the first peak. Indeed, molecules in the first shell also have a first solvation shell, and thus induce a second peak (for second nearest neighbors) in the rdf. This property is thus solvent induced and cannot be recovered by the (ideal) approximation of the excess (solvent-solvent) term. To conclude, a model as simple as an ideal fluid perturbed by a Lennard-Jones solute is sufficient to model the first solvation shell of a liquid neon.

Students are also asked to evaluate the number of neighbors in the first solvation shell of a given Ne. This can be done by integrating the radial distribution function from 0 to the end of the experimental first peak to count the number of molecules in this volume. Both experimental and cDFT radial distribution functions lead to a number of molecules that is close to 12, which corresponds to a compact packing between neon atoms.

At this point, it is a good thing to emphasis to students that the extraction of particular physical data can be easier with some computational techniques than with others. For instance, the generation of density maps such as in Figure 2, is completely straightforward from cDFT as it is a direct output of the theory. It can be interesting to discuss with them how they would compute the same quantities by explicit MD or MC simulation.
It is then the right moment to highlight the simplicity of the model and the crudeness of setting the excess term to zero. More complex systems can be studied with more sophisticated cDFT. State of the art cDFT requires approximately ten minutes to get a radial distribution function in quantitative agreement with Molecular Dynamics or Monte Carlo simulations that require 1000 CPU-hours.\cite{levesque_solvation_2012}

\begin{center}
\begin{figure}
\includegraphics[width=0.45\textwidth]{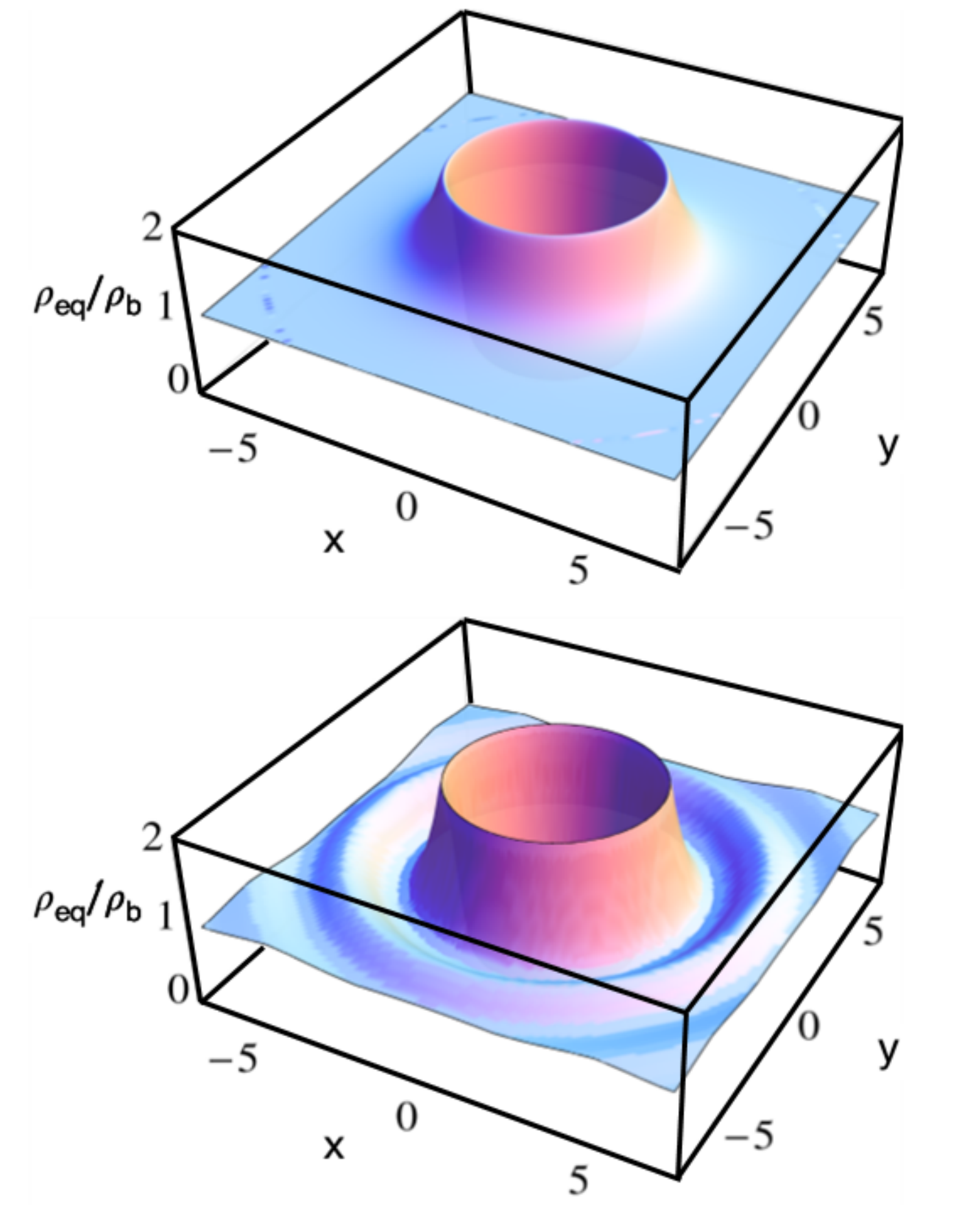}
\caption{On top, the density map predicted by cDFT for Ne in the approximation
described in the text. At bottom, the density map reconstructed from
the experimental radial distribution function. A single neon atom
is at coordinates $\left(x,y\right)=\left(0,0\right)$. \label{fig:On-top-is}}
\end{figure}

\par\end{center}

\section{Conclusions} 

It is worth to introduce classical density functional theory during a statistical mechanics course: it shows undergraduate students that liquid state theories exist and may be alternatives to the time-consuming and widely used Molecular Dynamics and Monte Carlo simulations. In order to illustrate how a cDFT code works, an in-silico experiment has been proposed and applied to a simple model system: liquid neon.
This experiment consisted in writing a Mathematica program that minimizes a functional both numerically and analytically. This little program, given in Supporting Information, can be written in 2 hours. With a simple model and their own code, students are able to reproduce the main features of the radial distribution function of liquid neon: The position and height of the first solvation shell. Even if this program is written for a simple physical problem, it introduces numerical minimization and some computational aspects of major interest. Last but not least, this activity has helped students to make a first contact with computational chemistry and state of the art theoretical research.

\bibliography{JCE}

\end{document}